\documentstyle[12pt,epsf]{article}

\begin{document}

\newcommand{\nc}{\newcommand}

\nc{\Ref}[1]{(\ref{#1})}
\nc{\nn}{\nonumber}
\nc{\el}{{\cal L}}
\nc{\D}{{\cal D}}
\def\boxit#1{\vbox{\hrule\hbox{\vrule\kern3pt
\vbox{\kern3pt#1\kern3pt}\kern3pt\vrule}\hrule}}

\renewcommand{\thefootnote}{\fnsymbol{footnote}}
\setcounter{footnote}{0}
\begin{titlepage}

\def\thefootnote{\fnsymbol{footnote}}

\begin{center}

\hfill IASSNS-HEP-00/22\\
\hfill YCTP-P3-00\\
\hfill hep-th/0003282\\
\hfill March, 2000\\
%\hfill \today

\vskip .5in

{\Large \bf Quantum Inconsistency of Einstein Supergravity}

\vskip .45in

{\large
  Jonathan A.\ Bagger,$^{a}$
  Takeo Moroi,$^{b}$\footnote
  {Address after April, 2000: Department of Physics, Tohoku
  University, Sendai 980-8578, Japan}
  and Erich Poppitz$^{c}$
}

\vskip 0.2in

{\em ${}^{a}$
  Department of Physics and Astronomy,
  The Johns Hopkins University\\
  Baltimore, MD 21218, USA}

\vskip 0.1in

{\em ${}^{b}$
  School of Natural Sciences,
  Institute for Advanced Study\\
  Princeton, NJ 08540, USA}

\vskip 0.1in

{\em ${}^{c}$
  Physics Department,
  Yale University\\
  New Haven, CT 06520, USA}

\end{center}

\vskip .4in

\begin{abstract}

We show that $N=1$, $D=4$ Einstein-frame supergravity is
inconsistent at one loop because of an anomaly in
local supersymmetry transformations.  A Jacobian
must be added to the Einstein-frame Lagrangian
to cancel this anomaly.  We show how the
Jacobian arises from the super-Weyl field redefinition that
takes the superspace Lagrangian to the Einstein frame.  We
present an
explicit example which demonstrates that the Jacobian is
necessary for one-loop scattering amplitudes to be frame
independent.

\end{abstract}

\end{titlepage}

\renewcommand{\thepage}{\arabic{page}}
\setcounter{page}{1}

\renewcommand{\thefootnote}{\#\arabic{footnote}}
\setcounter{footnote}{0}

\renewcommand{\theequation}{\thesection.\arabic{equation}}

\section{Introduction}
\setcounter{equation}{0}

The component Lagrangian of matter-coupled supergravity can
be derived from a superspace formulation or a tensor calculus
\cite{CFGP, WB}.  Both approaches inevitably lead to a component
theory in which the gravitational action is of a generalized
Brans-Dicke form,
\begin{equation}
e^{-1}{\cal L}\ =\ -\frac{1}{2}\,e^{-K/3}\,{\cal R}\ +\ \cdots\ .
\label{noneinstein}
\end{equation}
In this expression, $e$ is the vielbein determinant, ${\cal R}$
is the curvature scalar and $K$ is the K\"ahler potential, a
function of the scalar fields $A^*,A$.

The Lagrangian \Ref{noneinstein} leads to a kinetic mixing between
the graviton and the scalar fields. In addition, the kinetic terms
of the scalar fields appear in non-K\"ahler form.  It is therefore
convenient and customary to carry out a field-dependent Weyl rescaling
of the metric to bring the Lagrangian into canonical Einstein
form.  In supergravity, this Weyl rescaling must also be accompanied
by a chiral rotation of the fermions.  As we shall see, this rotation
gives rise to an anomalous Jacobian in the supergravity action.

The field redefinitions needed to go to this ``Einstein frame"
are usually performed in terms of component fields \cite{CFGP,
WB}.  This obscures the symmetries of the theory and complicates
the study of anomalies and their consequences. Therefore,
in this paper we will use the superspace approach of \cite{WB} to
study anomalies in supergravity theories.  We will show
that:
\begin{enumerate}
\item{The Einstein-frame field redefinitions can be carried out
directly in superspace through a super-Weyl transformation of the
vielbein.  The corresponding component-field Lagrangian gives
rise to ordinary Einstein gravity.}
\item{Local supersymmetry transformations in the Einstein frame
involve chiral rotations of the fermions.  They are anomalous
at one loop.}
\item{The local supersymmetry anomaly is cancelled by a Jacobian
that arises from the transition to the Einstein frame.  This
Jacobian is necessary to ensure the quantum consistency of
Einstein-frame supergravity.}
\item{The anomalous Jacobian can have important physical
consequences.   For example, it is necessary to ensure the
quantum equivalence of scattering amplitudes computed in
different frames.}
\end{enumerate}

This paper is organized as follows.  In Section~\ref{sec:SW}
we define super-Weyl transformations and derive the corresponding
Jacobians.  In Section~\ref{sec:Einstein} we study the transition
to the Einstein frame.  We show that one-loop supergravity invariance
of the Einstein-frame Lagrangian requires that a certain superspace
Jacobian be added to the bare Lagrangian.  In Section~\ref{sec:example}
we present an explicit example which illustrates the physical importance
of this Jacobian.  We summarize our results in Section~\ref{sec:summary}.

\section{Super-Weyl Transformations}
\setcounter{equation}{0}
\label{sec:SW}

\subsection{Classical Level}

In this section we study super-Weyl transformations in classical
and quantum supergravity.  These
transformations will play an important role throughout this
paper.

In what follows we use the notation and conventions of \cite{WB}.
We take the matter-coupled supergravity Lagrangian to be of the
form
\begin{eqnarray}
{\cal L} (X)
&=&
\int d^2 \Theta\ 2{\cal E}\ \Bigg[ \frac{3}{8}
\left( \bar{\D}^2 - 8 R \right) \exp\left\{ -\frac{1}{3}
\left(
K(\Phi^\dagger, \Phi) + \Gamma(\Phi^\dagger, \Phi, V) \right)
\right\}
\nonumber \\ &&
\ +\ \frac{1}{4} H_{ab}(\Phi) W^{(a)} W^{(b)}
\ + \ P(\Phi) \Bigg]\ +\ {\rm h.c.}\ ,
\label{L_SUGRA}
\end{eqnarray}
where $X = (\Phi, V, {E_M}^A)$ denotes the set of fields in
the supergravity Lagrangian, $\Phi$ and $V$ are the chiral
and vector superfields, and $E_M{}^A$  is the supervielbein.
In this expression, $K$ is the K\"ahler potential, $P$ the
superpotential, and $W^{(a)}$ the field strength superfield,
where $(a)$ is the index for the adjoint
representation.\footnote{In this paper, we sometimes omit
the spin index $\alpha$.  Therefore $W^{(a)}W^{(b)}$ should
be understood as $W^{(a)\alpha} W^{(b)}_\alpha$.}  In addition,
$H_{ab}$ is the gauge kinetic function, and $\Gamma$ is the
gauge counterterm which renders the Lagrangian gauge invariant.

In the superspace formalism, supergravity transformations are
given by translations in superspace.  Chiral and vector
superfields transform as follows,
\begin{eqnarray}
\label{susytransforms1}
\delta_{\rm SUSY} \Phi\ =\ -\xi^A \D_A \Phi\ ,~~~
\delta_{\rm SUSY} V\ =\ - \xi^A \D_A V\ ,
\end{eqnarray}
while the vielbein transforms as
\begin{eqnarray}
\label{susytransforms2}
\delta_{\rm SUSY} E_M{}^A\ =\ - \D_M \xi^A\ -\ \xi^B T_{BM}{}^A\ .
\end{eqnarray}
In these expressions, the $\D_A$ are covariant derivatives and the
$T_{BM}{}^A$ are the superspace torsion.  The superspace formalism
ensures that the Lagrangian (\ref{L_SUGRA}) is invariant,
\begin{eqnarray}
{\cal L} (X+\delta_{\rm SUSY}X)\ = \ {\cal L} (X)\ ,
\end{eqnarray}
up to a total derivative,
under the supersymmetry transformations \Ref{susytransforms1} and
\Ref{susytransforms2}.

Super-Weyl transformations are defined as rescalings of
the superspace vielbein that preserve the torsion
constraints \cite{HT}.\footnote{It is important
to distinguish between
super-Weyl and super-Weyl-K\"ahler transformations. The
latter are symmetries of the classical supergravity
Lagrangian.  A super-Weyl-K\"ahler transformation is a
super-Weyl transformation, with chiral superfield parameter
$\Sigma$, combined with a redefinition of the K\"ahler potential
and superpotential, $K\rightarrow K+6\Sigma+6\Sigma^\dagger$,
$P\rightarrow\exp(-6\Sigma)P$.}
They are parameterized by a chiral superfield $\Sigma$.  If
we denote the super-Weyl transformed $X$ field as $\hat{X}$,
we can write
\begin{eqnarray}
X\ =\ \hat{X}\ +\ \delta_{\rm SW} X\ ,
\label{wzredefinition}
\end{eqnarray}
where $\delta_{\rm SW} X$ is the super-Weyl variation of $X$.

To linear order in $\Sigma$, the super-Weyl transformations are
given by
\begin{eqnarray}
\delta_{\rm SW} {\cal E} &=& 6\Sigma\, {\cal E}\ +\
\frac{\partial}{\partial\Theta^\alpha}
\left( S^\alpha {\cal E} \right)
\label{dE(SW)} \nn\\
\delta_{\rm SW} \Phi &=&
- S^\alpha \frac{\partial}{\partial\Theta^\alpha} \Phi
\label{dPhi(SW)} \nn\\
\delta_{\rm SW} \left(\bar{\D}^2 - 8 R\right) U &=&
-\left( \bar{\D}^2 - 8 R\right)
\left(  4 \Sigma  - 2  \Sigma^\dagger \right) U
\ -\ S^\alpha \frac{\partial}{\partial\Theta^\alpha}
\left( \bar{\D}^2 - 8 R\right) U
\label{dU(SW)} \nn\\
\delta_{\rm SW} W_{\alpha} &=&
- 3 \Sigma\, W_{\alpha}
\ -\ S^\beta \frac{\partial}{\partial\Theta^\beta} W_{\alpha}\ ,
\label{dW(SW)}
\end{eqnarray}
where $U$ is any real vector superfield of Weyl
weight zero, and $S^\alpha$ is defined by
\begin{eqnarray}
\label{swz}
S^\alpha\ =\ \Theta^\alpha (2 \Sigma^\dagger - \Sigma)\vert
\ +\ \Theta^2 (\D^\alpha \Sigma)\vert\ .
\end{eqnarray}
The bar $\vert$ denotes the $\theta=\bar\theta=0$ component of
the superfield.

Note that
super-Weyl transformations also induce chiral rotations of the component
fermions.  Taking the appropriate components of Eq.~(\ref{dW(SW)}) and
re-exponentiating, we find
\begin{eqnarray}
\label{fermionswz}
\chi\ =\ \exp( \Sigma\vert - 2 \Sigma^\dagger\vert )\, \hat{\chi}\ ,
\qquad
\lambda\ =\ \exp(- 3\Sigma\vert)\, \hat{\lambda}\ ,
\end{eqnarray}
where $\chi$ and $\lambda$ are the fermions in the chiral and gaugino
multiplets, respectively.

In general, super-Weyl transformations are not symmetries of
the classical supergravity Lagrangian.  Indeed, substituting
the transformed variables \Ref{dW(SW)} into the Lagrangian
(\ref{L_SUGRA}), all derivative terms cancel. Nevertheless, a
nontrivial $\Sigma$ dependence remains.  At the classical level,
the Lagrangian (\ref{L_SUGRA}) becomes
\begin{eqnarray}
\label{WZlagr}
{\cal L} (\hat{X}+\delta_{\rm SW}X) &=&
\int d^2 \Theta\ 2 \hat{\cal E}\ \Bigg[ \frac{3}{8}
\left( \hat{\bar{\cal D}}^2 - 8 \hat{R} \right)
\exp\left\{ -\frac{1}{3}
\left( \hat{K} - 6\Sigma - 6\Sigma^\dagger + \hat{\Gamma}
\right) \right\}
\nonumber \\ &&
\ +\ \frac{1}{4} \hat{H}_{ab} \hat{W}^{(a)} \hat{W}^{(b)}
\ +\  \exp(6\Sigma) \hat{P} \Bigg]\ + \ {\rm h.c.}\ ,
\end{eqnarray}
where the hatted objects are evaluated using the Weyl-transformed
fields.  The K\"ahler and superpotential are different,
so the Lagrangian (\ref{L_SUGRA}) is not invariant.

\subsection{Quantum Level}

We are now ready to discuss the anomalous Jacobian associated
with a given super-Weyl transformation.  A proper framework is
provided by the one-particle-irreducible (1PI) effective
Lagrangian ${\cal L}_{\rm 1PI}$, defined by
\begin{eqnarray}
\int d^4x\  {\cal L}_{\rm 1PI} (X_{\rm C})
\ =\ -i \log\left[\ \int [dX_{\rm Q}]
\ \exp\left(i\int d^4x\ {\cal L}_{\rm bare}
(X_{\rm C}+X_{\rm Q}) \right)\right]\ .
\end{eqnarray}
Here ${\cal L}_{\rm bare}$ is the bare Lagrangian, and $X_{\rm C}$
and $X_{\rm Q}$ are classical and quantum parts of the $X$ field,
respectively.

In general, super-Weyl transformations are anomalous;
they  have a mixed super-Weyl-gauge anomaly.\footnote{It
also has a mixed super-Weyl-gravity anomaly.  We ignore
the gravity anomaly here.}\footnote{For a discussion
of supergravity anomalies in the compensator formalism,
see \cite{DWG}.}
Anomalies generate a set of non-local terms in the 1PI
effective action \cite{nonlocal}.
For the case at hand, the one-loop anomaly-induced terms are
\cite{CO, BMP},
\begin{equation}
\Delta {\cal L} \ =\ - \frac{1}{256\pi^2} \int d^2 \Theta
\ 2{\cal{E}}\ W^{(a)} W^{(a)}\
\frac{1}{\Box} \left( \bar{\cal D}^2 - 8R \right)
\left[ 4(T_R - 3 T_G) R^\dagger - \frac{1}{3} T_R {\cal D}^2 K
\right] \ +\ {\rm h.c.}\ ,
\label{ovrutterm}
\end{equation}
where we have omitted a term from the sigma-model anomaly that
is irrelevant for our discussion.  In Eq.~(\ref{ovrutterm}), $T_G$
is the Dynkin index of the adjoint representation, normalized to
$N$ for $SU(N)$, and $T_R$ is the Dynkin index associated with the
matter fields.  A sum over all matter representations is understood.
The first term, which contains the $R^\dagger$ superfield, arises
from the superconformal anomaly. It is proportional to the beta
function coefficient, $b_0 = 3 T_G - T_R$. The second term expresses
the K\"ahler anomaly \cite{CO, BMP}.

The variation of $\Delta {\cal L}$ can be computed by considering
a super-Weyl transformation with superfield parameter $\Sigma$.
Under such a transformation, the superfield $R$ changes as follows,
\begin{eqnarray}
\label{deltar}
\delta_{\rm SW} R\ = \
- 2 (2 \Sigma - \Sigma^\dagger)\, R \ -\ \frac{1}{4}\, {\bar\D}^2
\Sigma^\dagger\ .
\end{eqnarray}
This induces a shift by $\Sigma$ in the $R^\dagger$ term in
Eq.~(\ref{ovrutterm}).   A second shift comes from replacing
$K$ by $\hat{K}-6\Sigma-6\Sigma^\dagger$.  These two shifts
induce the following change in $\Delta{\cal L}$,
\begin{equation}
\Delta {\cal L} \ \rightarrow\ \Delta {\cal L}
\ +\ {\cal L}_{\rm J}\ ,
\end{equation}
where
\begin{eqnarray}
\label{jacobian}
{\cal L}_{\rm J}\ =\  \frac{1}{16\pi^2}
(3T_R - 3T_G) \int d^2 \Theta \ 2 {\cal E}\ \Sigma\,
W^{(a)} W^{(a)}
\ +\ {\rm h.c.}
\label{L_J}
\end{eqnarray}
The Lagrangian ${\cal L}_{\rm J}$ can be interpreted as
the superspace Jacobian that arises from the super-Weyl
transformation (\ref{wzredefinition}).  (Note that the imaginary
part of $\Sigma\vert$ corresponds to the Jacobian from the anomalous
$U(1)_R$ transformation.)

The nonvanishing Jacobian implies that
the functional measure is not invariant. It transforms as
follows under an arbitrary super-Weyl transformation:
\begin{eqnarray}
[d\Phi] [dV]\ =
\ [d(\hat{\Phi} +\delta_{\rm SW}\Phi )] [d(\hat{V}+
\delta_{\rm SW}V)]
\ =\ [d\hat{\Phi}] [d\hat{V}]
\exp \left[ i\int d^4x\ {\cal L}_{\rm J} \right]\ ,
\end{eqnarray}
The super-Weyl-rescaled 1PI Lagrangian is then
\begin{eqnarray}
\exp\left[i\int d^4x {\cal L}_{\rm 1PI} \right] &=&
\int[d(\hat{X}+\delta_{\rm SW}X)]
\ \exp\left[i\int d^4x
\ {\cal L}_{\rm bare} (\hat{X}+\delta_{\rm SW}X) \right]
\nonumber \\[2mm]
&=&
\int[d\hat{X}]
\ \exp\left[i\int d^4x
\left\{
{\cal L}_{\rm bare} (\hat{X})|_{
\hat{K}\rightarrow \hat{K}-6\Sigma - 6\Sigma^\dagger,
\ \hat{P}\rightarrow e^{6\Sigma}\hat{P}}
+ {\cal L}_{\rm J} \right\} \right]
\nonumber \\[2mm]
&=&
\int[d\hat{X}]
\ \exp\left[i\int d^4x\ \hat{\cal L}_{\rm bare} (\hat{X}) \right]\ ,
\end{eqnarray}
where
\begin{eqnarray}
\hat{\cal L}_{\rm bare} (\hat{X})\ \equiv
\ {\cal L}_{\rm bare} (\hat{X}+\delta_{\rm SW}X)
\ +\ {\cal L}_{\rm J}\ = \
{\cal L}_{\rm bare} (\hat{X})|_{
\hat{K}\rightarrow \hat{K}-6\Sigma - 6\Sigma^\dagger,
\ \hat{P}\rightarrow e^{6\Sigma}\hat{P}}
\ +\ {\cal L}_{\rm J}\ .
\end{eqnarray}
In these expressions, $\hat{\cal L}_{\rm bare}$ is the bare
Lagrangian for the quantum theory with super-Weyl-rescaled
variable $\hat{X}$.  The bare Lagrangian does not
contain the anomaly term $\Delta {\cal L}$, which
arises from integrating out the massless quantum fields.
It does, however, contain the Jacobian ${\cal L}_{\rm J}$.
As we will see, ${\cal L}_{\rm J}$ is important for ensuring
the quantum consistency of supergravity in the Einstein
frame.

\section{Einstein Supergravity}
\setcounter{equation}{0}
\label{sec:Einstein}

\subsection{The Einstein Frame}

In this section, we find the field-dependent Weyl
rescaling that takes the ``supergravity frame''
Lagrangian of Eq.~(\ref{L_SUGRA})
\begin{equation}
e^{-1}{\cal L}\ =\ -\frac{1}{2}\,e^{-K/3}\,{\cal R}\ +\ \cdots
\label{noneinstein2}
\end{equation}
into the Einstein frame.
In the literature, this rescaling has traditionally been done
in terms of component fields~\cite{CFGP,WB}.  Here we perform
the transformation in superspace.  This allows us to keep
better track of the symmetries of the theory.

The relevant superfield rescaling is, as we will see below, a
super-Weyl transformation with transformation parameter
$\Sigma_{\rm E}$,
\begin{eqnarray}
{\cal L}_{\rm E} (X) &=&
\int d^2 \Theta\ 2 {\cal E}\ \Bigg[ \frac{3}{8}
\left( \bar{\cal D}^2 - 8 R \right)
\exp\left\{ -\frac{1}{3}
\left( K - 6\Sigma_{\rm E} - 6\Sigma_{\rm E}^\dagger + \Gamma
\right) \right\}
\nonumber \\ &&
\ +\ \frac{1}{4} H_{ab} W^{(a)} W^{(b)}
\ +\  \exp(6\Sigma_{\rm E}) P \Bigg]\ +\ {\rm h.c.}\ ,
\label{L_E}
\end{eqnarray}
where we have omitted all ``hats'' in the above equation.  (Here
and hereafter, all quantities should be understood as being
defined in the frame obtained after the super-Weyl transformation,
unless specified otherwise.)

The parameter $\Sigma_{\rm E}$ can be found by demanding that
$K - 6\Sigma_{\rm E} - 6 \Sigma_{\rm E}^\dagger$ have no lowest,
$\Theta$ and $\Theta^2$ components since this combination
appears in the exponent
of the first term in Eq.~(\ref{L_E}).  To see, for example,
why $(K-6\Sigma_{\rm E}-6\Sigma_{\rm E}^\dagger)\vert$
must vanish, note that
Eq.~(\ref{noneinstein2}) involves the lowest component of
$e^{-K/3}$.  If the lowest component of this term is scaled
to 1, the factor $e^{-K/3}$ is absent and gravity is canonically
normalized.  The other two conditions lead to a canonical
kinetic term for the gravitino, and to canonical K\"ahler
kinetic terms for the matter multiplets.

The conditions on $\Sigma_{\rm E}$ are, therefore,
\begin{eqnarray}
\label{wzgaugeconditions}
K\vert\ =\ 6 \Sigma_{\rm E}\vert +  6 \Sigma_{\rm E}^\dagger\vert\ ,~~~
(\D_\alpha K)\vert\ =\ 6 (\D_\alpha \Sigma_{\rm E})\vert\ ,~~~
(\D^2 K)\vert\ =\ 6 (\D^2 \Sigma_{\rm E})\vert\ ,
\end{eqnarray}
or the vanishing of the lowest, $\Theta$ and $\Theta^2$
components of $K-6\Sigma_{\rm E}-6\Sigma_{\rm E}^\dagger$.  The Einstein frame
conditions (\ref{wzgaugeconditions}) almost completely determine
the parameter $\Sigma_{\rm E}$,
\begin{eqnarray}
\Sigma_{\rm E}\ =\ A_\Sigma\ +\ \sqrt{2} \Theta \chi_\Sigma
\ +\ \Theta^2 F_\Sigma\ ,
\label{SigmaE}
\end{eqnarray}
with
\begin{eqnarray}
A_\Sigma\ =\ \frac{1}{12} K + i\phi\ ,~~~
\chi_\Sigma\ =\ \frac{1}{6} K_i \chi^i\ ,~~~
F_\Sigma\ =\ \frac{1}{6} K_i F^i
- \frac{1}{12} K_{ij} \chi^i \chi^j\ ,
\label{Sigma_comp}
\end{eqnarray}
where the subscript $i$ on $K$ denotes the derivative with
respect to $A^i$ ($K_i\equiv \partial K/\partial A^i$), and
$F^i$ is the highest component of $\Phi^i$.  Note that
the conditions
\Ref{wzgaugeconditions} do not completely fix $\Sigma_{\rm E}$;
the imaginary part $\phi$ of its lowest component is left
undetermined.
Also note that $\Gamma$ does not contribute
to the conditions (\ref{wzgaugeconditions}): we assume that
a field redefinition is performed so that the matter fields
are in the Wess-Zumino gauge, where $\Gamma$ has no lowest,
$\Theta$ and $\Theta^2$ components.

With the above $\Sigma_{\rm E}$, it is not hard to find the component
Lagrangian.  If we substitute Eq.~(\ref{SigmaE}) into (\ref{L_E}),
all terms leading to non-Einstein gravity vanish.
The rest of the component Lagrangian can be readily evaluated
and gives the well-known Einstein supergravity Lagrangian with
canonical kinetic terms.   The complete expression
for the Lagrangian is given in \cite{CFGP, WB}.  We
have seen that the component Lagrangian can be directly
obtained from superspace, without any extra Weyl
rescalings of the component fields.

Let us now show that we can safely set the field $\phi$
to zero.  This field appears in the kinetic terms
of the matter and gauge fermions,
\begin{eqnarray}
\label{fermionkineticwz}
e^{-1} {\cal L}_{\rm kin} &=&
- i K_{ij^*} \bar{{\chi}}^{j^*} \bar\sigma^a
\left[ \partial_a + \frac{i}{6} b_a
- \frac{1}{6}
\left( K_k \partial_a A^k
- K_{j^*} \partial_a A^{j^*}
- 12 i \partial_a \phi \right) \right] \chi^i
\nonumber \\  &&
- i \bar{\lambda} \bar\sigma^a
\left(\partial_a - \frac{i}{2} b_a \right) \lambda\ .
\end{eqnarray}
(We omit terms with spin, sigma-model, and gauge connections
because they are not relevant for our discussion.) The field
$\phi$ also appears in the solution to the equations of motion
for the auxiliary field $b_a$,
\begin{eqnarray}
\label{bsolution}
b_a\ =\ \frac{i}{2}
\left( K_j \partial_a A^j
- K_{j^*} \partial_a A^{j^*}
- 12 i \partial_a \phi \right)
\ +\ \frac{1}{4} K_{ij^*}
\chi^i \sigma_a \bar{\chi}^{j^*}\ +\ \cdots\ .
\end{eqnarray}
(For the complete expression for $b_a$, see
Appendix~\ref{app:Jacobian}.)  In addition, it appears
in the superpotential terms in the Einstein frame,
\begin{eqnarray}
\label{phaseW}
e^{-1} {\cal L}_{\rm Yukawa}\ =
\ - \frac{1}{2} \exp \left(\frac{1}{2}K + 6 i \phi \right)
P_{ij} \chi^i \chi^j\ +\ {\rm h.c.}\ +\ \cdots\ .
\end{eqnarray}
(The complete expression for the Yukawa terms can be found in
\cite{CFGP, WB}.)

Upon inspection of Eqs.~(\ref{fermionkineticwz}), \Ref{bsolution}
and \Ref{phaseW}, one can check that the classical component
Lagrangian can be made independent of $\phi$ by redefining
$\chi\rightarrow e^{-3i\phi}\chi$ and $\lambda \rightarrow
e^{3i\phi}\lambda$.  In fact, the $\phi$ dependence also cancels
at the quantum level.  The field redefinitions used to go to
the Einstein frame are $\chi\rightarrow e^{-{\rm Re}\Sigma_{\rm E}\vert
+3i\phi}\chi$ and $\lambda\rightarrow e^{-3{\rm Re}\Sigma_{\rm E}\vert
-3i\phi}\lambda$.  The Jacobian from these transformations
exactly cancels the Jacobian from the redefinitions used to eliminate
$\phi$ from the component Lagrangian.  Therefore the
field $\phi$ is unphysical, and we can safely set it to
zero.

\subsection{Supersymmetry Transformations in the Einstein Frame}

In this subsection, we discuss the invariance of the classical
supergravity Lagrangian in the Einstein frame. Then, in the next
subsection, we consider quantum effects, and in particular,
anomalies.

It is simple to see that the Einstein-frame Lagrangian is not invariant
under the supersymmetry transformations (\ref{susytransforms1}).
Under a supersymmetry transformation, the K\"ahler potential transforms
as $\delta_{\rm SUSY}K=-\xi^A {\cal D}_A K$, in which case $\Sigma_{\rm E}$
transforms to $\Sigma_{\rm E}'$,
\begin{eqnarray}
\Sigma_{\rm E}'\ \equiv\ \Sigma_{\rm E}\,|_{\,K \rightarrow K+\delta_{\rm
SUSY}K}
\ =\ \Sigma_{\rm E}\ -\ \xi^A {\cal D}_A \Sigma_{\rm E}\ -\ \Sigma_\xi\ ,
\label{Sigma'}
\end{eqnarray}
where
\begin{eqnarray}
\label{compensatingsw}
\Sigma_\xi\ =\ - \frac{1}{6\sqrt{2}}
\left( \xi \chi^i K_i - \bar\xi \bar\chi^{i^*} K_{i^*} \right)
\ +\ {\cal O} (\Theta) \ +\  {\cal O} (\Theta^2)
\ \equiv\
i\phi_\xi\ +\ {\cal O} (\Theta)\ +\ {\cal O} (\Theta^2)\ .
\label{Sigma_xi}
\end{eqnarray}
The second term on the RHS of Eq.~(\ref{Sigma'}) is
the supersymmetry transformation of a normal chiral superfield.
The $\Sigma_\xi$ term is an additional super-Weyl
transformation under which the action is not invariant.

To see this explicitly, consider the supersymmetry transformation
of the Einstein-frame Lagrangian,
\begin{eqnarray}
{\cal L}_{\rm E}(X+\delta_{\rm SUSY}X) &=&
\int d^2 \Theta\ 2{\cal E}\ \Bigg[ \frac{3}{8}
\left( \bar{{\cal D}}^2 - 8 R \right)
\exp\left\{ -\frac{1}{3}
\left( K - 6(\Sigma_{\rm E}-\Sigma_\xi)
- 6(\Sigma_{\rm E}^\dagger-\Sigma_\xi^\dagger) + \Gamma
\right) \right\}
\nonumber \\ &&
\ +\ \frac{1}{4} H_{ab} W^{(a)} W^{(b)}
\ +\ \exp\left\{ 6(\Sigma_{\rm E} - \Sigma_\xi) \right\} P \Bigg]
\ +\ {\rm h.c.}
\label{L'_E}
\end{eqnarray}
Since $\Sigma_\xi$ is nonvanishing, the
Lagrangian is not invariant under the
supersymmetry transformation (\ref{susytransforms1}).

This lack of invariance stems from the fact that $\Sigma_\xi$
takes the Lagrangian out of the Einstein frame.  Invariance
can be restored by returning to the Einstein frame through
a compensating super-Weyl transformation.  This is similar
to gauge invariance in globally supersymmetric gauge theories.
There, a supersymmetry transformation in the Wess-Zumino gauge
must be supplemented by a superfield gauge transformation
to restore the Wess-Zumino gauge condition.
It is instructive to consider this case in some detail because
of the close analogy to supergravity \cite{KL}.  To that end,
we review the supersymmetry transformations of globally
supersymmetric gauge theories in Appendix~\ref{app:SYM}.

For the case at hand, the compensating super-Weyl transformation
has parameter $\Sigma_\xi$.  In the Einstein frame, therefore,
we {\it define} a supersymmetry transformation to include a
frame-restoring super-Weyl transformation with parameter
$\Sigma_\xi$:
\begin{eqnarray}
\delta_\xi\ \equiv\ \delta_{\rm SUSY}\ +\ \delta_{\rm SW}\ .
\label{del_SUSY(WZ)}
\end{eqnarray}
Under such a transformation, chiral and vector superfields
transform as follows,
\begin{eqnarray}
\delta_\xi \Phi\ =\ - \xi^A \D_A \Phi\ +\ \delta_{\rm SW} \Phi\ ,~~~
\delta_\xi V\ =\ - \xi^A \D_A V\ +\ \delta_{\rm SW} V\ ,
\label{dPhi&dV}
\end{eqnarray}
and analogously for the vielbein.
In these expressions, the first terms on the RHS are the
original supersymmetry transformations; the second
are the compensating super-Weyl transformations
with parameter $\Sigma_\xi$.
These transformations eliminate the $\Sigma_\xi$ in
Eq.~(\ref{L'_E}) and restore the classical
invariance of the action,
\begin{eqnarray}
{\cal L}_{\rm E} (X+\delta_\xi X)\ =\ {\cal L}_{\rm E} (X)\ .
\end{eqnarray}
The transformation properties of the individual
component fields can be derived by expanding
Eq.~(\ref{dPhi&dV}).  We have checked that they agree
with the transformations given in \cite{CFGP, WB} after
eliminating the auxiliary fields.

\subsection{Quantum Consistency in the Einstein Frame}

We are now ready to discuss anomalies in the supersymmetry
transformations (\ref{del_SUSY(WZ)}).  As we have seen,
these transformations include frame-restoring super-Weyl
field rescalings that induce chiral rotations on the matter
fermions,
\begin{eqnarray}
\label{chiralrotn}
\delta_\xi \chi \ =\ \cdots\ +\ 3i\phi_\xi\, \chi\ ,~~~
\delta_\xi \lambda\ =\ \cdots\ -\ 3i\phi_\xi\, \lambda\ ,
\end{eqnarray}
where $i\phi_\xi = \Sigma_\xi \vert$.  At the quantum level,
these transformations are anomalous, so they should give
rise to an anomalous variation of the 1PI Lagrangian,
\begin{eqnarray}
\delta_\xi {\cal L}_{\rm 1PI}\ =
\ e\left[ \frac{1}{16\pi^2}\, (3T_R - 3T_G)\,
\phi_\xi\, F^{(a)}_{mn} \tilde{F}^{mn(a)}\ +\ \cdots \right]\ ,
\label{XtraJac}
\end{eqnarray}
If nothing were to cancel this variation, local supersymmetry
in the Einstein frame would be anomalous.
In what follows, we will show that the full 1PI effective
action is, in fact, invariant.  The variation \Ref{XtraJac}
is cancelled by the variation of the Jacobian \Ref{L_J} that
arises in passing to the Einstein frame.

In the Einstein frame, the complete 1PI effective Lagrangian
is of the following form,
\begin{eqnarray}
{\cal L}_{\rm 1PI}\ = \
{\cal L}_{\rm E}\ +\ \Delta {\cal L}\ +\ {\cal L}_{\rm J}\ .
\label{L_1PI(E)}
\end{eqnarray}
The first term is the classical part of the Einstein-frame
Lagrangian, the second is the non-local term induced by
anomalies, and the third is the Jacobian \Ref{L_J}.
The first term is invariant under the local supersymmetry
transformation (\ref{del_SUSY(WZ)}), as discussed in the previous
subsection.  The second and third terms are not.  Under the
supersymmetry transformation (\ref{del_SUSY(WZ)}), the
nonvanishing variation of $\Delta {\cal L}$ expresses the
anomaly associated with the frame-restoring super-Weyl
transformation.\footnote{If
we make the argument before integrating out the light fields,
$\Delta {\cal L}$ does not exist.  In
this case the anomaly arises as a change of the functional
measure of the path integral.}
If this variation were the only change of the Lagrangian,
supersymmetry would be explicitly broken by anomalies at the
quantum level.

Fortunately, however, it is not.
There is also ${\cal L}_{\rm J}$, the Jacobian that arises in
the Einstein frame.  This term is {\it not} invariant under
(\ref{del_SUSY(WZ)}).  From Eq.~(\ref{Sigma'}), we have
\begin{eqnarray}
\delta_\xi \Sigma_{\rm E}\ =\ - \xi^A{\cal D}_A\Sigma_{\rm E}
\ -\ \Sigma_\xi\ .
\label{dSigma}
\end{eqnarray}
The first term on the RHS is the supersymmetry transformation
of a normal chiral superfield.  The second is a super-Weyl
transformation of $\Sigma$.  This gives
\begin{eqnarray}
\delta_\xi {\cal L}_{\rm J} &=&
- \frac{1}{16\pi^2} (3T_R - 3T_G)
\int d^2\Theta\ 2{\cal E}\ \Sigma_\xi W^{(a)} W^{(a)}
\ +\ {\rm h.c.}
\nonumber \\[2mm] &=&
e \left[ - \frac{1}{16\pi^2}\, (3T_R - 3T_G)\,
\,\phi_\xi\, F^{(a)}_{mn} \tilde{F}^{mn(a)}\ +\ \cdots \right]\ .
\label{dL_J}
\end{eqnarray}
Equation \Ref{dL_J} exactly cancels the variation (\ref{XtraJac}) and
restores supersymmetry invariance in the Einstein frame.

Thus we have seen that the Einstein-frame 1PI effective Lagrangian
is invariant under local supersymmetry transformations -- {\it
provided} the Jacobian (\ref{L_J}) is added to the bare Lagrangian.
Otherwise, local supersymmetry is explicitly broken at the quantum
level because of the anomaly associated with the frame-restoring
super-Weyl transformations.  The Jacobian (\ref{L_J}) is essential
for the consistency of the quantum theory.  The component expression
for the Jacobian is given in Appendix~\ref{app:Jacobian}.

\section{Physical Implications of the Jacobian: An Example}
\setcounter{equation}{0}
\label{sec:example}

In this section, we show that the anomalous Jacobian $\el_{\rm
J}$ has physical consequences.  We illustrate this with
an example of a scattering amplitude which requires the Jacobian
to give a frame-independent result.

In what follows we consider a model with a no-scale K\"ahler
potential of the form
\begin{eqnarray}
\label{examplekahler}
K \ =\ -3 \log \left(1 - {1 \over 3} \Phi^\dagger \Phi \right)\ .
\end{eqnarray}
This K\"ahler potential (\ref{examplekahler}) is chosen for
simplicity; upon substitution into the superspace Lagrangian
(\ref{L_SUGRA}), it gives rise to canonically normalized
scalars with a conformal coupling to gravity,
\begin{eqnarray}
{\cal L}_{\rm SUGRA}\ = \
\sqrt{-g}\, \left[\, -\frac{1}{2} \left(1 - \frac{1}{3} A^*A \right){\cal R}\
-\ g^{mn} \partial_m A^* \partial_n A\ +\ \cdots \right]\ ,
\label{R|A|^2}
\end{eqnarray}
where we used the metric $g_{mn}$ instead of a vielbein.  In
Eq.~(\ref{R|A|^2}), the subscript SUGRA indicates that this is
the supergravity-frame Lagrangian.  As discussed in the
previous sections, we can use a super-Weyl transformation
to pass to the Einstein frame.

\begin{figure}
\centerline{\epsfxsize=0.8\textwidth\epsfbox{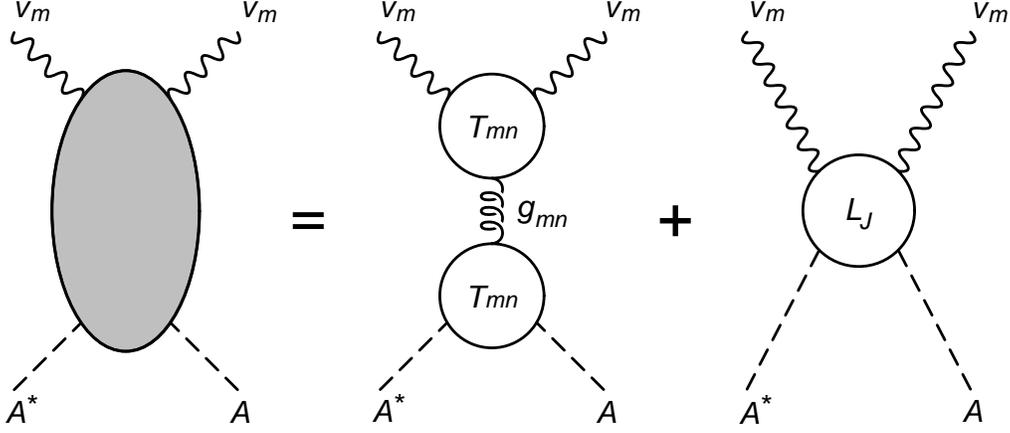}}
\caption{Schematic picture of the amplitude calculation.}
\label{fig:feyndiag}
\end{figure}

In this model, we study the scattering process $A^*A\rightarrow
v_mv_m$ via a graviton exchange, where $v_m$ is a gauge boson
and $A$ is a massless neutral scalar field.  For simplicity, we
assume that the gauge theory is pure supersymmetric Yang
Mills.\footnote{If
the matter fields had gauge quantum numbers, the following
discussion would still hold, provided we replace $3T_G$ by
$3T_G-T_R$.}

The amplitude of this process can be written as follows (see
Fig.~\ref{fig:feyndiag})
\begin{eqnarray}
i{\cal M} (A^*A \rightarrow v_mv_m) &=&
\frac{i}{2}\ \langle v_m (\epsilon_1, p'_1) v_m (\epsilon_2, p'_2)
| \,T^{kl}\, | 0 \rangle
\ \times\ i \Delta_{kl,mn}\nn\\
&&\ \times\
\frac{i}{2}\, \langle 0 |\, T^{mn}\, | A^* (p_1) A (p_2) \rangle
\nn\\[2mm]
&& + \
\langle v_m (\epsilon_1, p'_1) v_m (\epsilon_2, p'_2) |
\,i {\cal L}_{\rm J}\, | A^* (p_1) A (p_2) \rangle\ ,
\label{amplitude}
\end{eqnarray}
where the $p_i$ denote the momenta of the scalars, while
$\epsilon_i$ and $p'_i$ denote the polarizations and
momenta of the gauge bosons.  Here, $T_{mn}$ is the
energy momentum tensor,
\begin{eqnarray}
 T_{mn}\ =\ {2\over \sqrt{- g}}\, {\delta{\cal L}\over \delta g^{mn}}\ ,
\end{eqnarray}
while $\Delta_{kl,mn}$ is the graviton propagator,
given by
\begin{eqnarray}
\label{graviton}
\Delta_{mn, kl} (q)\ =\ - \frac{2}{q^2}
( \eta_{km} \eta_{ln} + \eta_{kn} \eta_{lm}
- \eta_{kl} \eta_{mn} )\ ,
\end{eqnarray}
where $\eta_{mn}$ is the flat-space metric.  We omit the part
that depends on the gauge parameter because it does not contribute
to the amplitude of interest.

For the scalars, the matrix element of the energy-momentum tensor
is
\begin{eqnarray}
\langle 0 |\, T^{mn}\, | A^* (p_1) A (p_2) \rangle
\ = \ p_1^m p_2^n\ +\ p_1^n p_2^m\ -\ (p_1 p_2) \eta^{mn}
\ +\ \frac{\zeta}{3} (q^2 \eta^{mn} - q^m q^n)\ ,
\label{Tscalar}
\end{eqnarray}
where $q\equiv p_1+p_2$, and $\zeta$ is a coefficient proportional
to the coupling of the scalars to the scalar curvature ${\cal R}$;
$\zeta =1$ in the supergravity frame and $\zeta =0$ in the Einstein
frame.  Note that $\zeta =1$ corresponds to conformally
coupled scalars; this is easy to see upon computing the trace of
Eq.~(\ref{Tscalar}) with $q^2=2p_1p_2$.

Substituting Eqs.~(\ref{graviton}) and \Ref{Tscalar} into
Eq.~(\ref{amplitude}), we obtain the amplitude
\begin{eqnarray}
\label{amplitude2}
{\cal M} &=& \frac{2}{q^2} p_{1k} p_{2l}
\langle v_m v_m |\, T^{kl}\, | 0 \rangle
\ +\  \frac{\zeta}{6} \langle v_m v_m | T^k_k | 0 \rangle
+ \langle v_m v_m | \,{\cal L}_{\rm J}\, | A^* A \rangle
\nonumber \\ &\equiv&
{\cal M}_0
\ +\  \frac{\zeta}{6} \langle v_m v_m |\, T^k_k\, | 0 \rangle
\ + \ \langle v_m v_m | \,{\cal L}_{\rm J}\, | A^* A \rangle\ ,
\end{eqnarray}
where ${\cal M}_0$ is frame independent.  At the classical
level, with ${\cal L}_{\rm J}$ absent, the amplitudes in
both frames are identical because the gauge boson energy
momentum tensor is traceless.

At the quantum level, the frame independence is more subtle.
First, the gauge boson energy
momentum tensor is no longer traceless.   Second, there
is a Jacobian in the Einstein frame, but not the supergravity
frame.  To see what happens, let us consider the
supergravity frame.  We take $\zeta=1$ and use the trace
anomaly relation
\begin{eqnarray}
T^k_k\  =\  \frac{3}{32\pi^2} \,T_G\, F_{mn}^{(a)} F^{mn(a)}\ ,
\label{Tkk}
\end{eqnarray}
to find
\begin{eqnarray}
{\cal M}_{\rm SUGRA}\  =\  {\cal M}_0\  +
\ \frac{3}{192\pi^2} \, T_G\,
\langle v_m v_m | \,F_{mn}^{(a)} F^{mn(a)}\, | 0 \rangle\ .
\end{eqnarray}
In the Einstein frame, where $\zeta=0$, there is no
contribution from the $T^k_k$ term.  However, in this
frame there is the super-Weyl Jacobian,
\begin{eqnarray}
{\cal L}_{\rm J} &=&  \frac{1}{16\pi^2} (-3T_G)
\int d^2 \Theta\  2 {\cal E}\  \Sigma_{\rm E}\, W^{(a)} W^{(a)}\  +\  {\rm
h.c.}\nn\\
&=& \frac{3}{192\pi^2} \,T_G\, A^*A\, F_{mn}^{(a)} F^{mn(a)}\ +\ \cdots\ ,
\label{L_J(SYM)}
\end{eqnarray}
where we have used $\Sigma_{\rm E}\vert =\frac{1}{12}A^*A+\cdots$.
With this Jacobian, the Einstein-frame matrix element is
\begin{eqnarray}
{\cal M}_{\rm E}\ =\ {\cal M}_0 \ + \
\frac{3}{192\pi^2} \,T_G\,
\langle v_m v_m |\, F_{mn}^{(a)} F^{mn(a)}\, | 0 \rangle\ ,
\end{eqnarray}
which is in complete agreement with the amplitude in the supergravity
frame.  The results in the two frames are identical
because of the super-Weyl Jacobian.

\section{Summary}
\setcounter{equation}{0}
\label{sec:summary}

In this paper, we studied the quantum consistency of the supergravity
Lagrangian.  We used a superspace approach in which the supergravity
Lagrangian does not automatically give canonically normalized
Einstein gravity.  In the literature, Einstein
gravity is recovered after a redefinition of the component
fields.  In this paper, we showed that the field redefinition is, in
fact, a super-Weyl transformation, and we demonstrated a systematic
way to do the field redefinition in superspace.  This approach
provides us with a clear understanding of supersymmetry
transformations  in the Einstein frame.

Supersymmetry transformations in the Einstein frame must preserve
the Einstein frame condition, so they differ from the original
supersymmetry transformations defined in the supergravity frame.
We showed that the Einstein-frame supersymmetry
transformations are ordinary supersymmetry transformations,
$\delta_{\rm SUSY}$, combined with compensating super-Weyl
transformations, $\delta_{\rm SW}$, which are necessary to maintain
the Einstein frame condition.

The compensating super-Weyl transformations are, at the quantum level,
anomalous. Because of this fact, one must be careful when studying
the supersymmetry invariance of the quantum effective action.  In
this paper we emphasized that the super-Weyl transformation used
to pass to the Einstein frame is anomalous.  It
gives rise to an anomalous
Jacobian that must be included in the bare Einstein-frame Lagrangian.
The 1PI Einstein-frame Lagrangian is supersymmetric because the
variation of this Jacobian precisely cancels the anomaly arising
from the frame-restoring super-Weyl transformation.  If the
Jacobian were omitted, the 1PI Lagrangian would not be invariant
under Einstein-frame supersymmetry transformations.  Consistency
demands that the Jacobian be included in the bare Einstein-frame
Lagrangian.

\section*{Acknowledgments}

The work of J.B.\ is supported by the U.S.\ National Science
Foundation, grant NSF-PHY-9970781.  T.M.\ is
supported by U.S.\ National Science Foundation under grant
NSF-PHY-9513835, and by the Marvin L.\ Goldberger Membership.  E.P.\
is supported by the Department of Energy, contract DOE
DE-FG0292ER-40704.

\appendix

\section{Globally Supersymmetric Gauge Theories}
\setcounter{equation}{0}
\label{app:SYM}

In this Appendix, we show how the supersymmetry transformations
are defined in globally supersymmetric gauge theories.  In particular,
we demonstrate how the Wess-Zumino gauge condition is
maintained after supersymmetry transformations.  We will see
that there is a close analogy between supersymmetry transformations
in globally supersymmetric gauge theories and supergravity
transformations in the Einstein frame.

Consider a globally supersymmetric gauge theory, with the Lagrangian
\begin{eqnarray}
\label{globallagr}
{\cal L}\ =\ {\cal L}_{\rm gauge}\ +\ \int d^4 \theta\ \Phi^\dagger
e^{V} \Phi\ ,
\label{L_SYM}
\end{eqnarray}
where ${\cal L}_{\rm gauge}$ is the kinetic term for the gauge
multiplet.  This Lagrangian is invariant under the superspace
supersymmetry transformations
\begin{eqnarray}
\label{globalsusy}
\delta_{\rm SUSY} V\ =\  \xi^A \partial_A V\ , ~~~
\delta_{\rm SUSY} \Phi\ =\ \xi^A \partial_A \Phi\ ,
\end{eqnarray}
where $\xi^a = -i\xi\sigma^a\bar\theta +i\theta\sigma^a\bar\xi$
and $\xi^\alpha$ is the supersymmetry transformation parameter.
The component transformations can be determined by expanding
(\ref{globalsusy}) in powers of $\theta$.

The Lagrangian (\ref{L_SYM}) contains the lower components of $V$,
which are gauge degrees of freedom.
Usually, it is convenient to use a Lagrangian in which these lower
components are eliminated by a gauge transformation.  This
``frame'' is often called ``the Wess-Zumino gauge;'' it is obtained
by the following field redefinitions:
\begin{eqnarray}
V_{\rm WZ}\ =\ V  - \Lambda - \Lambda^\dagger\ ,~~~
\Phi_{\rm WZ}\ =\  e^{\Lambda} \Phi\ ,
\end{eqnarray}
where $\Lambda$ is a chiral superfield.  The gauge parameter
$\Lambda$ is chosen so that all the unphysical fields are
eliminated from the Lagrangian.  The conditions are
\begin{eqnarray}
V\vert\ =\ \Lambda\vert\ +\ \Lambda^\dagger\vert\ ,~~~
(D_\alpha V)\vert\ =\ (D_\alpha\Lambda)\vert\ ,~~~
(D^2 V)\vert\ =\ (D^2 \Lambda)\vert\ ,
\end{eqnarray}
so $\Lambda$ in the chiral basis is simply
\begin{eqnarray}
\label{lambda}
\Lambda\ =\ \frac{1}{2}C\ +\ i \phi\ +\ i \theta \chi\ +
\ \frac{i}{2} \theta^2 (M + i N)\ ,
\end{eqnarray}
where $C$, $\chi$, $M+iN$ are the lowest, $\theta$ and $\theta^2$
components of the vector superfield $V$, respectively.  Note that
the field $\phi$ is not determined by the Wess-Zumino
gauge conditions; it is the parameter of an ordinary gauge
transformation.  In terms of the new variables, the Lagrangian
becomes
\begin{eqnarray}
{\cal L}_{\rm WZ} &=& {\cal L}_{\rm gauge}\ +\
\int d^4\theta\
\Phi_{\rm WZ}^\dagger\, e^{V-\Lambda-\Lambda^\dagger} \Phi_{\rm WZ}\nn\\
&=& {\cal L}_{\rm gauge}\ +\ \int d^4\theta\ \Phi_{\rm WZ}^\dagger
\,e^{V_{\rm WZ}}\, \Phi_{\rm WZ}\ .
\label{L_WZ}
\end{eqnarray}
The Wess-Zumino gauge Lagrangian ${\cal L}_{\rm WZ}$ contains
only the physical fields.

The Wess-Zumino gauge conditions, however, are not preserved by
the supersymmetry transformations (\ref{globalsusy}).   They must
be supplemented by compensating superfield gauge transformations,
\begin{eqnarray}
\label{globalsusywz}
\delta_\xi V_{\rm WZ}\ =\ \xi^A \partial_A V_{\rm WZ}\
+\ \Lambda_\xi\ ,~~~
\delta_\xi \Phi_{\rm WZ}\ =\ \xi^A \partial_A \Phi_{\rm WZ}
\ +\  \Lambda_\xi \Phi_{\rm WZ}\ ,
\end{eqnarray}
where, in the chiral basis,
\begin{eqnarray}
\label{lambdaxi}
\Lambda_\xi\ =\  \theta \sigma^m \bar\xi
\,(-2 v_m + 2 \partial_m \phi)\ +\ \theta^2 \bar\xi \bar\lambda\ .
\end{eqnarray}
Here $v_m$ and $\lambda$ are the gauge boson and gaugino fields,
respectively.

The transformations (\ref{globalsusywz}) are combinations of the
original supersymmetry transformations (\ref{globalsusy})
and the frame-restoring gauge transformations.  They leave
invariant the Wess-Zumino-gauge Lagrangian.
Furthermore, if we expand
the LHS of (\ref{globalsusywz}) in powers of $\theta$, we
obtain the Wess-Zumino-gauge supersymmetry transformations
given, for example, in \cite{WB}.  Note that in Wess-Zumino
gauge, a theory with a gauge anomaly would also have a
supersymmetry anomaly because the transformations
(\ref{globalsusy}) contain ordinary gauge transformations.

Finally, we comment on the difference between
(\ref{lambdaxi}) and (\ref{compensatingsw}),
that is, on the different way we treat the imaginary
part of the lowest component of the compensating
transformation parameters.  In each case, the term
is not determined by
the Wess-Zumino gauge/Einstein frame conditions.  In
globally supersymmetric gauge theory, we choose not
to fix $\phi_\xi$ in $\Lambda_\xi\vert$;  it is the
degree for freedom associated
with an ordinary gauge transformation.  By contrast, in
supergravity, we completely fix it and
demand that the imaginary part of $\Sigma_{\rm E}\vert$ vanish.
The first term in (\ref{compensatingsw}) ensures that
the imaginary part of  $\Sigma_{\rm E}\vert$ does not reappear
in the Lagrangian after a supersymmetry transformation.

\section{Component Expression for the Jacobian}
\setcounter{equation}{0}
\label{app:Jacobian}

In this Appendix, we present the complete component expression for
the Jacobian that arises from the super-Weyl transformation required
to pass to the Einstein frame.

As we have seen in this paper, the bare Lagrangian in
Einstein-frame supergravity is given by
\begin{eqnarray}
\hat{\cal L}_{\rm bare}\ =\ {\cal L}_{\rm E}\ +\ {\cal L}_{\rm J}
\end{eqnarray}
where ${\cal L}_{\rm E}$ is the classical supergravity Lagrangian
whose component expression is given, for example, in \cite{CFGP, WB}.
${\cal L}_{\rm J}$ is the Jacobian.  At one-loop level,
${\cal L}_{\rm J}$ is given by
\begin{eqnarray}
{\cal L}_{\rm J} = \frac{1}{16\pi^2} (3T_R-3T_G)
\int d^2\Theta\ 2{\cal E}\ \Sigma_{\rm E}\, W^{(a)} W^{(a)}
\ +\ {\rm h.c.} \ ,
\label{L_Jac}
\end{eqnarray}
where the chiral superfield $\Sigma_{\rm E}$ is given in Eqs.~(\ref{SigmaE})
and \Ref{Sigma_comp} with $\phi=0$,
\begin{eqnarray}
\Sigma_{\rm E}\ =\ A_\Sigma\ +\ \sqrt{2} \Theta \chi_\Sigma
\ +\ \Theta^2 F_\Sigma \ ,
\end{eqnarray}
with
\begin{eqnarray}
A_\Sigma\ =\ \frac{1}{12} K\ ,~~~
\chi_\Sigma \ =\ \frac{1}{6} K_i \chi^i\ ,~~~
F_\Sigma\ =\ \frac{1}{6} K_i F^i
\ -\ \frac{1}{12} K_{ij} \chi^i \chi^j\ .
\end{eqnarray}
Expanding Eq.~(\ref{L_Jac}), we obtain
\begin{eqnarray}
e^{-1} {\cal L}_{\rm J} &=& \frac{1}{16\pi^2}\, (3T_R-3T_G)
\  \Bigg[
- A_\Sigma F_{mn}^{(a)} {F^{mn}}^{(a)}
\nonumber \\ &&
-\ 2i A_\Sigma \lambda^{(a)} \sigma^m
\left( {\cal D}_m \bar{\lambda}^{(a)}
- f^{abc} v^{(b)}_m \bar{\lambda}^{(c)}
+ \frac{i}{2} b_m \bar{\lambda}^{(a)} \right)
\nonumber \\ &&
-\ 2i A_\Sigma \bar{\lambda}^{(a)} \bar{\sigma}^m
\left( {\cal D}_m \lambda^{(a)}
- f^{abc} v^{(b)}_m \lambda^{(c)}
- \frac{i}{2} b_m \lambda^{(a)} \right)
\nonumber \\ &&
+\ 2 A_\Sigma {D_{\rm aux}}^{(a)} {D_{\rm aux}}^{(a)}
\nonumber \\[2mm] &&
+\ i A_\Sigma
\left(
\psi_m \sigma^{kl} \sigma^m \bar{\lambda}^{(a)}
+ \bar{\psi}_m \bar{\sigma}^{kl} \bar{\sigma}^m \lambda^{(a)}
\right)
\left( F_{kl}^{(a)} + {\hat{F}_{kl}}^{(a)} \right)
\nonumber \\ &&
-\ \sqrt{2} \left( \chi_\Sigma \sigma^{mn} \lambda^{(a)}
+ \bar{\chi}_\Sigma \bar{\sigma}^{mn} \bar{\lambda}^{(a)} \right)
F_{mn}^{(a)}
\nonumber \\ &&
+\ \sqrt{2}i \left(
\chi_\Sigma \sigma^{mn} \lambda^{(a)}
\psi_m \sigma_n \bar{\lambda}^{(a)}
+ \frac{1}{4} \bar{\psi}_m \bar{\sigma}^m \chi_\Sigma
\lambda^{(a)}\lambda^{(a)} \right)
\nonumber \\ &&
+\ \sqrt{2}i \left(
\bar{\chi}_\Sigma \bar{\sigma}^{mn} \bar{\lambda}^{(a)}
\bar{\psi}_m \bar{\sigma}_n \lambda^{(a)}
+ \frac{1}{4} \psi_m \sigma^m \bar{\chi}_\Sigma
\bar{\lambda}^{(a)} \bar{\lambda}^{(a)} \right)
\nonumber \\ &&
+\ \sqrt{2}i \left( \chi_\Sigma \lambda^{(a)}
- \bar{\chi}_\Sigma \bar{\lambda}^{(a)} \right)
{D_{\rm aux}}^{(a)}
\nonumber \\ &&
-\ F_\Sigma \lambda^{(a)}\lambda^{(a)}
\ -\ F^*_\Sigma \bar{\lambda}^{(a)} \bar{\lambda}^{(a)}
\Bigg]
\end{eqnarray}
where ${\hat{F}_{mn}}^{(a)}$ is the supercovariant field
strength,
\begin{eqnarray}
{\hat{F}_{mn}}^{(a)}\ =\ F_{mn}^{(a)}
\ -\ \frac{i}{2} \left(
\psi_m \sigma_n \bar{\lambda}^{(a)}
 + \bar{\psi}_m \bar{\sigma}_n \lambda^{(a)}
 - \psi_n \sigma_m \bar{\lambda}^{(a)}
 - \bar{\psi}_n \bar{\sigma}_m \lambda^{(a)} \right)\ .
\end{eqnarray}
For a detailed explanation of the notation, see \cite{WB}.

The full component expression is given by substituting the solutions
to the Einstein-frame auxiliary field equations of motion,
\begin{eqnarray}
F^i &=& (K^{-1})^{ij^*}
\left( -e^{K/2} D_{j^*} P^*\ +\ \frac{1}{2} K_{j^*kl} \chi^k \chi^l
\ +\ \frac{1}{4} \partial_{j^*} h_{(ab)}
\bar{\lambda}^{(a)} \bar{\lambda}^{(b)} \right)\nn
\\
{D_{\rm aux}}^{(a)} &=& -\, {h^{{\rm R}(ab)}}^{-1}
\left[ D^{(b)}
\ +\ \frac{i}{2\sqrt{2}} \left( \partial_i h_{(bc)} \chi^i \lambda^{(c)}
\ -\ \partial_{i^*} h_{(bc)} \bar{\chi}^i \bar{\lambda}^{(c)}
\right) \right]\nn
\\
b_m &=& \frac{i}{2} \left( K_i \tilde{\cal D}_m A^i
- K_{i^*} \tilde{\cal D}_m A^{*i} \right)
\ +\ \frac{1}{4} K_{ij^*} \chi^i \sigma_m \bar{\chi}^j
\nonumber \nn\\ &&
-\ \frac{3}{4} h_{(ab)}^{\rm R}
\lambda^{(a)} \sigma_m \bar{\lambda}^{(b)}
\ + \ i \left[ \frac{1}{2} \left( K_i X^{i(a)}
- K_{i^*} X^{*i(a)} \right) + i D^{(a)} \right] v^{(a)}_m\ ,
\end{eqnarray}
where $D_iP\equiv P_i+K_iP$, $X^{(a)}$ is the Killing vector,
$D^{(a)}$ is the Killing potential associated with $X^{(a)}$, and
$\tilde{\cal D}_mA^i\equiv {\cal D}_mA^i - v_m^{(a)} X^i_{(a)}$
\cite{WB}.

\end{document}